\documentclass{article}

\usepackage{PRIMEarxiv}

\usepackage[utf8]{inputenc} % allow utf-8 input
\usepackage[T1]{fontenc}    % use 8-bit T1 fonts
\usepackage{hyperref}       % hyperlinks
\usepackage{url}            % simple URL typesetting
\usepackage{booktabs}       % professional-quality tables
\usepackage{amsfonts,amsmath}       % blackboard math symbols
\usepackage{nicefrac}       % compact symbols for 1/2, etc.
\usepackage{microtype}      % microtypography
\usepackage{lipsum}
\usepackage{fancyhdr}       % header
\usepackage{graphicx}       % graphics
\graphicspath{{media/}}     % organize your images and other figures under media/ folder

\usepackage{bm}

\usepackage{algorithm}
\usepackage{algpseudocode}

\newcommand{\ra}{{\rm a}}

\usepackage{tikz}
\usepackage{pgfplots}
\usepackage{pgfplotstable}
\pgfplotsset{
compat=newest, % version compability
tick label style={font=\footnotesize}, % size of the tick labels
}

\title{PCWE for FSAI -- Derivation of scalar wave equations for fluid-structure-acoustics interaction of low Mach number flows
}

\author{
	\href{https://orcid.org/0000-0002-2148-6703}{\includegraphics[scale=0.06]{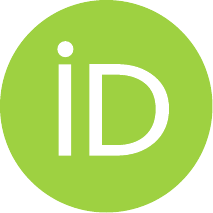}\hspace{1mm}Stefan Schoder} \\
    Institute of Fundamentals and Theory in Electrical Engineering (IGTE)\\
	Graz University of Technology\\
	8010 Graz, Austria \\
	\texttt{stefan.schoder@tugraz.at} \\
}

\begin{document}
\maketitle

\begin{abstract}
This paper presents a novel derivation of the perturbed convective wave equation by utilizing the instantaneous velocity field as the foundation for the convective operator. This approach holds particular significance in the context of modeling fluid-structure-acoustic interaction (FSAI) problems, such as those encountered in human phonation or systems involving fans with flexible blades. The derivation is explored through two distinct methodologies: one commencing from the linearized perturbed compressible equations (LPCE) and the other rooted in the acoustic perturbation equations of variant 2 (APE-2). The comparative analysis of these alternative derivations may contribute to a more profound understanding of the underlying physics involved in FSI scenarios and aeroacoustics. 
\end{abstract}

% keywords can be removed
\keywords{Aeroacoustics \and Fluid dynamics \and Acoustics \and Helmholtz's decomposition \and Flow solver \and APE \and LPCE \and FSI \and FEM}

\section{Introduction}
\label{sec:Intro} 
Aeroacoustic analogies \cite{Lighthill1952} compute noise radiation efficiently but the obtained fluctuating field converges only in steady flow regions to the acoustic field~\cite{schoder2019hybrid,kaltenbacher2021physical}. %Additionally,
%an ambiguity of the acoustic analogies arises
%since the sources depend on the solution of the
%equations \cite{schoder2019diss}. 
As recognized by Phillips \cite{PhillipsEQ} and Lilley \cite{LilleyEQ}, the source terms responsible for mean flow-acoustics interactions should be part
of a convective wave operator.
Therefore, an aeroacoustic approach based on a
systematic decomposition of the field properties emerged. This circumvents that sources depend on the acoustic solution and provides a rigorous definition of acoustics. Ribner \cite{Ribner1962} decomposed the fluctuating pressure in a pseudo pressure and an acoustic pressure part $p' =
p^0 + p^\mathrm{a} $. Hardin and Pope \cite{HP1994} formulated their
viscous/acoustic splitting technique expansion about the
incompressible flow (EIF). The EIF
formulation was modified and applied to several examples \cite{Shen1999,Shen1999a,Slimon1999}. 
Ewert and Schröder \cite{Ewert2003} proposed a different technique leading to the acoustic perturbation equations (APE). Instead of filtering the flow field, the source terms of the
wave equations are filtered according to the characteristic properties of the acoustic modes. The acoustic modes are obtained from LEE. Recent advances of the theory led to \cite{ewert2021hydrodynamic}. Hüppe \cite{Hueppe} derived a computationally efficient reformulation of the APE-2 system and named it perturbed convective wave equation (PCWE). Since then, a number of low Mach number flow applications have been addressed by computational aeroacoustics using the PCWE model. This includes human phonation \cite{schoder2023implementation,schoder2020hybrid,schoder2021aeroacoustic,falk20213d,lasota2021impact,maurerlehner2021efficient,schoder2022error,valavsek2019application,lasota2023anisotropic,kraxberger2022machine}, HVAC systems \cite{tautz2019aeroacoustic} and fan noise \cite{schoder2020computational,tieghi2022machine,schoder2019conservative,schoder2021application,tieghi2023machine} using openCFS \cite{schoder2022opencfs,schoder2023opencfs}. The PCWE is valid for incompressible flows and near-field acoustics \cite{Maurerlehner2022}. Recently, a convective wave equation was proposed for investigating jet-noise and installation effects \cite{schoder2023acoustic,schoder2024correction,schoder2022aeroacoustic,schoder2022cpcwe,schoder2024aeroacoustic}.

Regarding the state-of-the-art, we derive a wave equation that considers relevant second-order fluctuation products omitted in the derivation of the APE-2 system to obtain a convective wave operator that depends on the instantaneous incompressible flow field. A similar derivation can be performed starting from the linearized perturbed compressible equations (LPCE) \cite{Seo2006}. Both equations are derived within this contribution and can be directly applied to fluid-structure-acoustic interaction problems of technical (like ducted fans with flexible blades at high rotational speeds \cite{schoder2023affordable}, fluid-structure interactions with their acoustic impact in hydraulic turbines \cite{lenarcic2015numerical}) and biological flows (like the fluid-structure-acoustic interactions reported in \cite{kraxberger2023alignment,nager2023investigation}). In summary, the correct modeling of human phonation lacks of a consistent definition of an aeroacoustic wave equation with the fluid-structure interaction process of the vocal fold motion, as indicated in \cite{dollinger2023overview}.

\section{Theory }

The theory is based on the incompressible Navier-Stocks equation for the fluid field modeled by a large eddy simulation and the Navier equations using linear-elastic material or hyper-elastic material for the solid mechanical field. The fluid structure interaction process is coupled by two coupling conditions of the fluid dynamic and solid mechanical field \cite{kaltenbacher2007numerical}. The first fluid-solid interaction boundary condition on the non-permeable interface $\Gamma_{\mathrm{fs}}$ is of an inhomogeneous Dirichlet type for the flow velocity $u$
\begin{equation}
    \bm{u}=\frac{\partial \bm{d}}{\partial t} \quad \text { on }\quad(0, T) \times \Gamma_{\mathrm{fs}}
\end{equation}
and the mechanical displacement $d$. The movement of the fluid-solid interface $\Gamma_{\mathrm{fs}}$ changes of the computational fluid domain, which needs special care during the computation. The second fluid-solid boundary condition for the solid is given by an inhomogeneous Neumann condition or flux condition, resulting in continuous normal stresses according to the stress tensor $\left[\boldsymbol{\sigma}\right]$ and the normal $\boldsymbol{n}$ along the interface $\Gamma_{\mathrm{fs}}$
\begin{equation}
\left[\boldsymbol{\sigma}_{\mathrm{s}}\right] \cdot \boldsymbol{n}=\left[\boldsymbol{\sigma}_{\mathrm{f}}\right] \cdot \boldsymbol{n} = \boldsymbol{f}_{\text {fluid }}=\underbrace{-p \mathcal{I} \cdot \boldsymbol{n}}_{\text {Pressure }}+\underbrace{\mu_{\mathrm{f}}\left(\nabla \boldsymbol{u}+(\nabla \boldsymbol{u})^T\right) \cdot \boldsymbol{n}}_{\text {Shear }} \quad \text { on }\quad(0, T) \times \Gamma_{\mathrm{fs}}.
\end{equation}
In the usual case, the feedback of the acoustics on flow and structural deformation is negligible and therefore not considered in the following discussion. However, as shown in \cite{kraxberger2023alignment,nager2023investigation} there are relevant cases for low Mach number flow applications where acoustic feedback is relevant. 

\subsection{Modified PCWE for FSAI}
Based on the preliminary work of the PCWE, we derive a convective wave equation based on the instantaneous incompressible velocity field. %As used during the derivation of the APE system, we first apply linearization and then distinguish between vortical and acoustical perturbations.
The PCWE can be derived efficiently from the APE-2 system \cite{Ewert2003} and the use of Helmholtz decomposition \cite{schoder2020helmholtz}
\begin{eqnarray}
p &=&p_0 + p'  = \bar p + p_{\mathrm ic} + p_\ra  \\ \rho &=& \rho_0 + \rho'\\
\bm u &=&\bm u_0 + \bm u' =  \bm u_0
+ \bm u_{\mathrm ic} + \bm u_{\mathrm a} = \bm u_0 + \nabla \times \bm A - \nabla \psi_\ra \,.
\end{eqnarray}
with the pressure $p$, the mean pressure $p_0$, the perturbation pressure $p'$, the incompressible perturbation pressure $ p_{\mathrm ic}$, the acoustic perturbation pressure $p_\ra$, the density $\rho$, the mean density $\rho_0$, the perturbation density $\rho'$, the velocity $\bm u$, the mean velocity $\bm u_0$, the perturbation  velocity $\bm u'$, the incompressible perturbation velocity $\bm u_{\mathrm ic} = \nabla \times \bm A$, the acoustic perturbation velocity $\bm u_{\mathrm a} = - \nabla \psi_\ra$ and the vector potential $\bm A$. We define the compressible acoustic field as irrotational field that can be modeled by the acoustic scalar potential $\psi_\ra$ as a gradient field of the acoustic particle velocity
%\begin{equation}
$\bm u_\ra = -\nabla \psi_\ra $.
We arrive from an general flow field at the following perturbation
equations
\begin{eqnarray}
\label{eq:APE1}
\frac{\partial p_\ra}{\partial t} - c_0^2 \frac{\partial \rho'}{\partial t} &=& - \frac{\partial p_{\mathrm ic}}{\partial t} \label{eq:p}\\[2mm]
\frac{\partial \rho'}{\partial t} + \nabla \cdot ( \rho_0 \bm u_{\mathrm a} + \bm u_0 \rho') &=& 0 \label{eq:rho} \\[2mm]
\frac{\partial \bm u_\ra}{\partial t} + \nabla \big(
\bm u_0 \cdot \bm u_\ra \big) + \nabla \frac{p'}{\rho_0} &=& \frac{1}{\rho_0} \nabla p_{\mathrm ic}\ \label{eq:momentum}
\end{eqnarray}
with the isentropic speed of sound $c_0$. Combining equation (\ref{eq:p}) and (\ref{eq:rho}) results in the intermediate result
\begin{equation}
    \frac{\partial p_\ra}{\partial t} - c_0^2 \nabla \cdot ( \rho_0 \bm u_{\mathrm a} + \bm u_0 \rho') = - \frac{\partial p_{\mathrm ic}}{\partial t} \, .
\end{equation}
Before, proceeding with the derivation, the previously discarded (during the derivation of the APE) product of the $\bm u_{\mathrm ic} \rho'$ is added again
\begin{equation}
    \frac{\partial p_\ra}{\partial t} - c_0^2 \nabla \cdot ( \rho_0 \bm u_{\mathrm a} + \bm u_0 \rho' + \bm u_{\mathrm ic} \rho') = - \frac{\partial p_{\mathrm ic}}{\partial t} \, .
\end{equation}
Now the fluctuating density can be eliminated by using the relation of the pressure and density of $p' = c_0^2\rho'$ and apply the divergence to the term. The divergence of the incompressible flow is zero and therefore, the following result is obtained
\begin{equation}
    \frac{\partial p_\ra}{\partial t} - c_0^2 \nabla \cdot ( \rho_0 \bm u_{\mathrm a}) + c_0^2  (\bm u_0 + \bm u_{\mathrm ic})  \cdot \nabla p' = - \frac{\partial p_{\mathrm ic}}{\partial t}  \, .
\end{equation}
Splitting the pressure into its fluctuating parts yields the following equation
\begin{equation}
    \frac{\partial p_\ra}{\partial t}  +  (\bm u_0 + \bm u_{\mathrm ic})  \cdot \nabla p_{\mathrm a} - c_0^2 \nabla \cdot ( \rho_0 \bm u_{\mathrm a}) = - \frac{\partial p_{\mathrm ic}}{\partial t}  - (\bm u_0 + \bm u_{\mathrm ic})  \cdot \nabla p_{\mathrm ic} \, .
\end{equation}
At this stage, the material derivative for this convective wave equation is defined by $\mathrm{D}/\mathrm{D} = \partial /\partial t  +  (\bm u_0 + \bm u_{\mathrm ic})  \cdot \nabla$ with $\bm u_0 + \bm u_{\mathrm ic}$ being the incompressible field of an incompressible flow simulation. This convective derivative is suitable for modeling acoustics on deformed grids conveniently and for fluid-structure interaction problems where one has no parametrization of the grid deformation before the simulation. In some cases, for instance with the use of an ALE reference frame, the original PCWE equation can be used \cite{kaltenbacher2017aiaa} but the original PCWE is not that convenient for the general FSAI case. This is the case because the complete time-history must be stored to compute the convective operator. Another benefit is that instantaneous coupling of the acoustic equations with the flow is possible. Using the definition of the material derivative, we arrive at
\begin{equation}
    \frac{\mathrm{D} p_\ra}{\mathrm{D} t}  - c_0^2 \nabla \cdot ( \rho_0 \bm u_{\mathrm a}) = - \frac{\mathrm{D} p_{\mathrm ic}}{\mathrm{D} t} \, . \label{eq:someResult}
\end{equation}
Rewriting equation (\ref{eq:momentum}) with the scalar potential and discarding the gradient yields the following intermediate result %Vertauschen von zeit und ort..theorem
\begin{equation}
    -\frac{\partial \psi_\ra}{\partial t} - 
\bm u_0 \cdot \nabla \psi_\ra  +  \frac{p'}{\rho_0} = \frac{1}{\rho_0} p_{\mathrm ic} \, .
\end{equation}
The second-order interaction $-\bm u_{\mathrm ic} \cdot \nabla \psi_\ra$ is again added to the equation 
\begin{equation}
    -\frac{\partial \psi_\ra}{\partial t} - 
(\bm u_0 + \bm u_{\mathrm ic}) \cdot \nabla \psi_\ra   = \frac{p_{\mathrm ic} - p'}{\rho_0} = - \frac{p_{\mathrm a}}{\rho_0}\, .
\end{equation}
Rewriting the left-hand side terms with the material derivative, we 
define the acoustic pressure
\begin{equation}
\label{eq:pressAPE1}
p_{\mathrm a} = \rho_0  \frac{\mathrm{D} \psi_\ra}{\mathrm{D} t} \, ,
\end{equation}
Finally, we insert the definition of the acoustic pressure and the definition of the acoustic potential into equation (\ref{eq:someResult}) and obtain the PCWE for FSAI applications
\begin{equation}
\label{eq:cPCWE}
\frac{1}{c_0^2} \, \, \frac{\mathrm{D}^2\psi_\ra}{\mathrm{D} t^2} - \Delta \psi_\ra =
- \frac{1}{\rho_0 c_0^2}\, \frac{\mathrm{D}  p_{\mathrm ic}}{\mathrm{D} t} \,.
\end{equation}
This convective wave equation describes acoustic sources
generated by incompressible FSAI applications and their wave propagation
through flowing media. In addition, instead of the original unknowns
$p_\ra$ and $\bm v_\ra$, we have just one scalar
$\psi_\ra$ unknown. An additional benefit of this modified PCWE is that the boundary conditions can be described consistently with the scalar potential and are independent of the aeroacoustic source generation. Additionally, to the equation inside the fluid \eqref{eq:cPCWE}, the following two boundary conditions have to be considered as well. Firstly, the continuity at a solid-fluid interface requires a coinciding normal component of the mechanical surface velocity of the solid with the normal component of the acoustic velocity of the fluid
\begin{equation}
\label{eq:BC1}
\boldsymbol{n} \cdot \frac{\partial \boldsymbol{d}}{\partial t}=-\boldsymbol{n} \cdot \nabla \psi_\ra \quad \text { on }\quad(0, T) \times \Gamma_{\mathrm{fs}} .
\end{equation}
Secondly, the acoustic fluid pressure causes on the surface a mechanical stress in the normal direction
\begin{equation}
\label{eq:BC2}
\left[\boldsymbol{\sigma}_{\mathrm{s}}\right] \cdot \boldsymbol{n}=-\boldsymbol{n} p_{\mathrm{a}}=-\boldsymbol{n} \rho_0 \frac{\mathrm{D} \psi_\ra}{\mathrm{D} t}\, .
\end{equation}

\subsection{Modified scalar variant of LPCE for FSAI}
Similar to the derivation of the modified PCWE, one can derive a convective wave for incompressible FSAI problems \cite{piepiorka2019numerical}. Based on the preliminary work of LPCE \cite{Seo2006}, we derive a convective wave equation using a scalar potential. The primary quantities of the LPCE are defined as
\begin{eqnarray}
p &=&P_{\mathrm ic} + p_\ra  \\ \rho &=& \rho_0 + \rho'\\
\bm u &=&\bm U_{\mathrm ic} + \bm u_{\mathrm a} = \nabla \times \bm B + \nabla \phi_\ra \,.
\end{eqnarray}
with the pressure $p$, the incompressible pressure $P_{\mathrm ic}$, the acoustic perturbation pressure $p_\ra$, the density $\rho$, the mean density $\rho_0$, the perturbation density $\rho'$, the velocity $\bm u$, the incompressible velocity $\bm U_{\mathrm ic} = \nabla \times \bm B$, the acoustic perturbation velocity $\bm u_{\mathrm a} = - \nabla \phi_\ra$ and the vector potential $\bm B$. We define the compressible acoustic field as irrotational field that can be modeled by the acoustic scalar potential $\phi_\ra$.
We arrive from a general compressible flow at the following perturbation
equations
\begin{eqnarray}
\label{eq:LPCE}
\frac{\partial \rho'}{\partial t} + \bm U_{\mathrm ic} \cdot \nabla \rho' +
\rho_0 \nabla \cdot \bm u_\ra &=& 0 \\[2mm]
\frac{\partial p_\ra}{\partial t} + \bm U_{\mathrm ic} \cdot \nabla p_\ra +
\gamma P_{\mathrm ic} \nabla \cdot \bm u_\ra + \bm u_\ra \cdot \nabla P_{\mathrm ic}   &=& - \frac{\mathrm{D}  P_{\mathrm ic}}{\mathrm{D} t} \label{eq:pressureLPCE} \\[2mm]
\frac{\partial \bm u_\ra}{\partial t} + \nabla \big(
\bm U_{\mathrm ic} \cdot \bm u_\ra \big) + \frac{1}{\rho_0} \nabla p_\ra &=& 0\ \label{eq:momentumLPCE}
\end{eqnarray}
with the ratio of specific heats $\gamma$. The linearized momentum equation (\ref{eq:momentumLPCE}) can be rewritten with the definition of the acoustic scalar potential and the elimination of the gradient to
\begin{equation}
    \rho_0 \frac{\partial \phi_\ra}{\partial t} + 
\bm U_{\mathrm ic} \cdot \nabla \phi_\ra  - p_\ra = 0
\end{equation}
and further yields the definition of the acoustic pressure
\begin{equation}
\label{eq:pressLPCE}
p_{\mathrm a} = \rho_0  \frac{\mathrm{D} \phi_\ra}{\mathrm{D} t} \, .
\end{equation}
Rewriting the pressure equation (\ref{eq:pressureLPCE}) using the acoustic potential yields the intermediate result
\begin{equation}
\label{eq:interLPCE}
\frac{1}{\rho_0} \frac{\mathrm{D} p_\ra}{\partial t} +
\frac{\gamma P_{\mathrm ic}}{\rho_0} \nabla \cdot \nabla \phi_\ra + \nabla \phi_\ra \cdot \nabla P_{\mathrm ic}   = - \frac{1}{\rho_0} \frac{\mathrm{D}  P_{\mathrm ic}}{\mathrm{D} t} \, .
\end{equation}
For an ideal gas, we can introduce the speed of sound as $c_0^2 = \frac{\gamma P_{\mathrm ic}}{\rho_0}$. Finally, equation (\label{eq:pressLPCE}) is inserted into equation (\label{eq:interLPCE}) and we obtain the following wave equation
\begin{equation}
\label{eq:cPCWE2}
\frac{1}{c_0^2} \, \, \frac{D^2\phi_\ra}{D t^2} + \frac{1}{\rho_0 c_0^2} \nabla \phi_\ra \cdot \nabla P_{\mathrm ic} - \Delta \phi_\ra =
- \frac{1}{\rho_0 c_0^2}\, \frac{D \Phi_\mathrm{p}}{D t} \,.
\end{equation}
This convective wave equation describes acoustic sources
generated by incompressible flow structures and their wave propagation
through flowing media. In addition, instead of the original unknowns
$p_\ra$ and $\bm v_\ra$, we have just one scalar
$\psi_\ra$ unknown. Additionally to the modified PCWE for FSAI, this equation has an additional interaction term on the left-hand side $\frac{1}{\rho_0 c_0^2} \nabla \phi_\ra \cdot \nabla P_{\mathrm ic}$ which will be investigated in the future. On the interface from the solid to the fluid, the first condition is identically to \eqref{eq:BC1} and a negligible back-coupling based on condition \eqref{eq:BC2}.

	\section{Conclusions and Outlook}
	
This short working paper presents a draft derivation of the PCWE for FSI applications and we are happy to receive feedback. 

In conclusion, we presented a first exploration of the extended version of the perturbed convective wave equation, employing the instantaneous velocity field as a fundamental element of the convective operator. The derived equations, originating from both linearized perturbed compressible equations (LPCE) and acoustic perturbation equations of variant 2 (APE-2), offer valuable equations for fluid-structure-acoustic interaction (FSAI) problems. The two derivations provide two perspectives, enhancing the understanding of aeroacoustic phenomena observed in applications. These findings not only contribute to the theoretical foundation of FSAI modeling but also hold practical implications for the accurate simulation and prediction of fluid acoustic responses in real-world scenarios.

\bibliographystyle{elsarticle-num}
\bibliography{mybibfile}

\end{document}